# Bounded Semantic Planning and Deterministic Compilation for Reliable Enterprise Text-to-SQL

Yi Ai

**Abstract**

Direct text-to-SQL asks a language model to do two jobs: interpret the business question and construct the complete relational query. In enterprise schemas, SQL can execute successfully while using the wrong relationship role or aggregation grain. We study an alternative placement of the stochastic boundary. A multi-turn planner grounds phrases and selects from question-specific governed options; graph traversal, role predicates, grain lowering, SQL construction, and deterministic checks are implemented in code. We evaluate this semantic path compilation (SPC) system against direct DDL-to-SQL generation on the ACME insurance benchmark. On a 38-question adjudicated comparison set with three runs per question, SPC was adjudicated correct on every run for 37 questions (97.4%), compared with 21 (55.3%) for the baseline. The paired discordance was 16 questions in favor of SPC and none in favor of the baseline (two-sided exact McNemar $p = 3.05 \times 10^{-5}$). SPC answered all 38 questions correctly at least once and produced one refusal and no adjudicated wrong-but-executed run across 114 run outcomes; the baseline produced 29 adjudicated wrong runs and seven additional judge-flagged data-only coincidences on the same set. A strict-equivalence sensitivity analysis increased the paired difference. Additional SPC runs with GPT-5.4 and Gemini-3.6-Flash showed similar question-level robustness, although their per-run verdict artifacts were not preserved. Six additional benchmark items are retained in an all-item analysis and documented separately by failure class. The study supports an end-to-end systems result, not a causal claim that compilation alone produced the gain, because SPC receives governed semantic artifacts that the DDL baseline does not.

## 1. Introduction

Generating valid SQL is only part of enterprise text-to-SQL. A system must also select the intended business interpretation, choose among relationally valid paths, bind role-sensitive relationships, and preserve grain across joins and aggregation. Errors in these decisions often produce executable SQL and plausible numbers. Parser validity and agreement on one database instance are therefore insufficient safeguards.

Semantic layers move some of this work into governed metadata. MetricFlow, for example, composes joins among modeled entities and generates SQL deterministically [dbt-join-logic; dbt-2026]. On dbt's original minimally modeled ACME configuration, its 2026 study reports 72.7% overall for the Semantic Layer and 64.5% for direct text-to-SQL; the Semantic Layer answered the configured in-scope questions but not those outside its modeled scope [dbt-2026]. Ontology-mediated systems such as ATHENA [Saha-2016] and ATHENA++ [Sen-2020] likewise construct ontology-level interpretations before translating them to SQL. These systems establish that semantic graph traversal and deterministic translation are not new.

This paper asks a narrower systems question: **what happens when an LLM is allowed to interpret the question but not to author joins, aggregation structure, or SQL?** Figure 1 shows the boundary we evaluate.

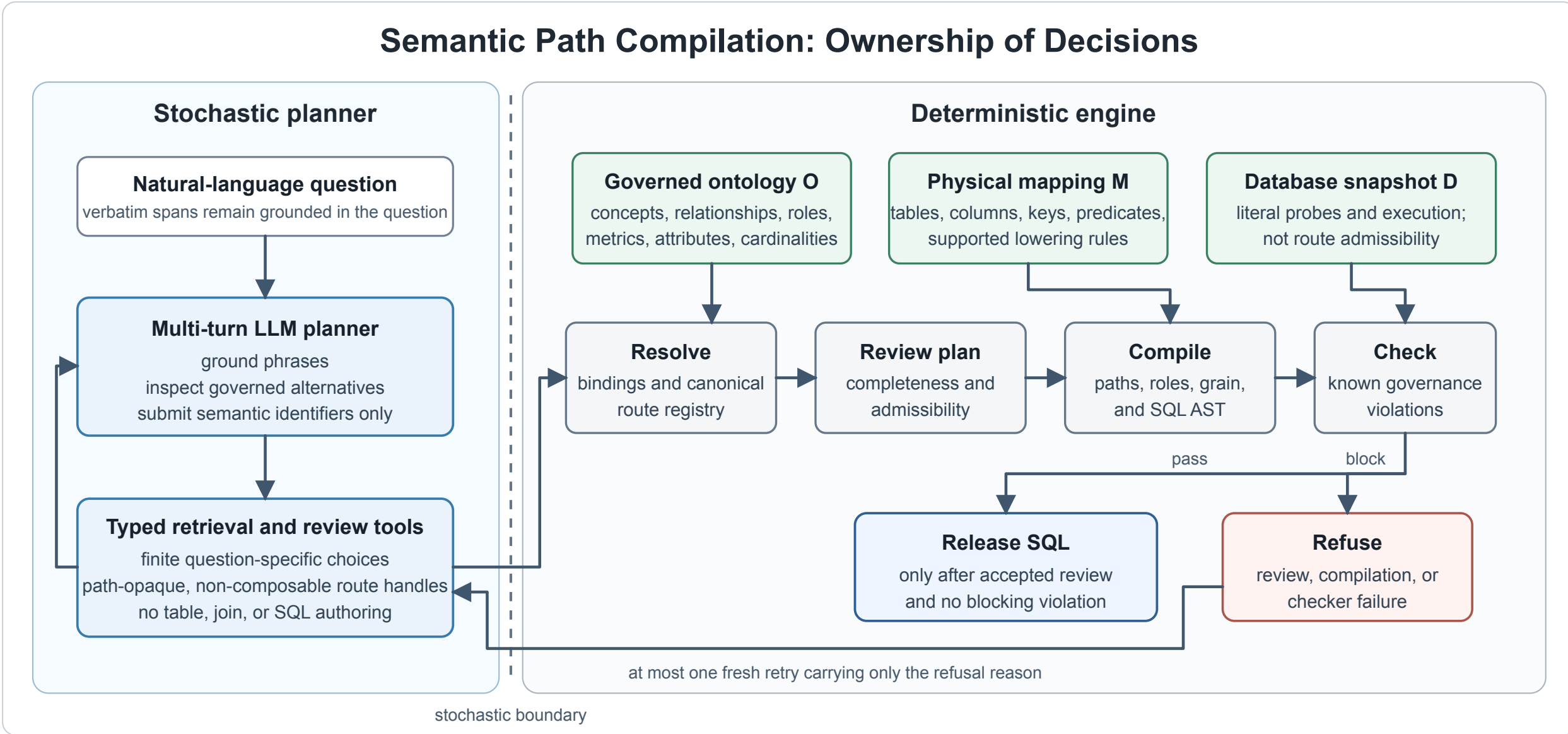

**Figure 1.** Semantic path compilation places the stochastic boundary after language grounding and governed selection. Typed retrieval tools expose a finite question-specific choice set; the deterministic engine reviews the submitted plan, resolves paths and roles, lowers grain, constructs SQL, and checks known governance conditions.

The distinction is operational. The model may quote spans from the question, select a subject, and choose path-opaque, non-composable identifiers returned by deterministic tools. It cannot name a table, column, join predicate, role predicate, or SQL fragment. The engine reviews the submitted plan again, compiles it, and checks it for known governance violations. A rejected attempt may be retried once in a fresh conversation carrying only the refusal reason.

**Contributions.**

1. **Bounded semantic planning.** A text-to-SQL architecture that confines the model to grounding and selection over dynamically retrieved, question-specific semantic alternatives (§4).
2. **Deterministic relational realization.** A compiler that converts the accepted semantic plan into complete SQL, including governed paths, role predicates, grouping, and supported grain transformations (§4).
3. **Multiplicity-focused evaluation.** Counterfactual database instances designed to separate grain-safe and grain-unsafe queries that coincide on the shipped fixture (§5).
4. **A paired reliability study.** Question-level repeated-run, refusal, and wrong-but-executed measurements against a direct DDL-to-SQL baseline, with an exact paired test and a strict-scoring sensitivity analysis (§6).

The scope is intentionally limited. The experiment compares complete systems on one domain, the compiler was developed while inspecting the benchmark, and the evaluation does not hold semantic knowledge constant between arms.

## 2. Background and Related Work

### 2.1 Ontology-mediated and semantic-layer querying

ATHENA maps natural language to ontology query language through ranked interpretation trees and then translates the ontology-level query to SQL [Saha-2016]. ATHENA++ extends this line to nested and aggregation queries [Sen-2020]. Ontop supports virtual ontology-based data access by rewriting formal SPARQL queries over OWL 2 QL ontologies and relational mappings into SQL [Ontop-2017]. These systems demonstrate both ontology-level path construction and deterministic database translation; our contribution is not either mechanism in isolation.

Commercial semantic layers expose a related contract. MetricFlow composes joins through modeled entities and generates SQL within the modeled semantic surface [dbt-join-logic]. dbt's ACME study provides an external reference on the same benchmark substrate [dbt-2026]. Rumiantsau and Fokeev [2604.25149] report gains from semantic context across three models, while Sequeda et al. [Sequeda] and Allemang and Sequeda [Allemang-2024] study knowledge-graph querying and ontology-based query checking on the benchmark lineage used here.

### 2.2 Intermediate representations and constrained generation

IRNet lowers SemQL, an intermediate representation generated by a model, into SQL [Guo-2019]. PICARD constrains autoregressive decoding by rejecting inadmissible continuations through incremental SQL parsing [Scholak-2021]. Kim et al. [2606.31041] compile a Semantic Model Query through a semantic layer. These approaches reduce the SQL search space or move SQL construction behind an IR. SPC differs in where it places the remaining model freedom: the planner may select only vocabulary and route handles returned for the current question, and the final SQL is a deterministic function of the accepted plan, ontology, physical mapping, and compiler version. This is a difference in system contract, not a claim that typed IRs or deterministic lowering are new.

### 2.3 Grain and aggregation correctness

Aggregation placement and summarizability are established database problems. Yan and Larson [Yan-1995] analyze eager aggregation around joins. Lenz and Shoshani [Lenz-1997] formalize summarizability conditions, and Hurtado and Mendelzon [Hurtado-2001] give reasoning procedures for heterogeneous multidimensional schemas. Grain Theory [Karayannidis-2026] provides a recent type-level account of grain propagation and fan traps, while leaving pipeline synthesis open. The present work implements a restricted compiler policy for the ACME ontology and measures it; it does not claim a new general theory of grain.

### 2.4 Evaluation and selective reliability

Single-database execution accuracy can accept semantically different queries that happen to agree on one instance. Zhong et al. [Zhong-2020] address this problem with distilled suites of database instances. Our counterfactual instances follow the same general principle but target role- and multiplicity-sensitive enterprise errors rather than claiming a new semantic-equivalence criterion.

Reliability work also treats abstention as a first-class outcome. TrustSQL [TrustSQL] penalizes incorrect answers more heavily than abstention, and ontology-based query checking reports explicit “I don’t know” outcomes [Allemang-2024]. SQLStructEval [2604.06736] measures structural variation among execution-correct text-to-SQL outputs. We therefore report answered correctly, refused, and wrong-but-executed separately, and evaluate stability at the question level rather than presenting only mean run accuracy.

## 3. Problem Definition

### 3.1 The enterprise analytical query task

```
Given:
  Q — a natural-language analytical question
  D — an enterprise relational database
  S — semantic metadata describing the business concepts,
      metrics, attributes and relationships of D

Produce:
  SQL q  such that  q(D)  answers Q according to the
  intended business semantics of S.
```

### 3.2 Relational decisions left to generation

Under direct generation, the model must decide the relationship path among several relationally valid routes, the business role where one physical relation carries several, the metric interpretation, the aggregation grain, the filters and their attachment points, and the SQL realization:

```
Q + schema context
  → generated relational plan and SQL
  → executable answer or failure
```

The risk studied here is not that generation is always wrong, but that relational decisions are implicit in sampled output and may be difficult to govern independently. SPC turns a subset of those decisions into explicit typed choices and deterministic transformations.

### 3.3 Outcomes

- **Task success:** the returned rows satisfy the question under the stated adjudication rule.
- **Solid success:** all three runs for a question succeed.
- **At-least-once success:** one or more of three runs succeed.

- **Refusal:** the system returns no SQL.
- **Wrong-but-executed:** SQL runs but the result is adjudicated incorrect.

Solid and at-least-once success are question-level summaries. Refusals and wrong-but-executed outcomes are retained separately because they have different operational consequences.

### 3.4 Research question

> On ACME, how do bounded semantic planning and deterministic relational realization compare with direct DDL-to-SQL generation in repeated-run task success, refusal, and wrong-but-executed outcomes?

This question is comparative rather than causal. The experiment does not separate the value of semantic knowledge from the mechanism that executes it.

### 3.5 The deterministic contract

Let $Q$ be a natural-language question, $\mathcal{O}$ the governed semantic model, $\mathcal{M}$ the physical mapping, and $D$ the database snapshot. The governed graph induced by the model and mapping is

$$\mathcal{G}(\mathcal{O}, \mathcal{M}) = (V, E),$$

where $V$ contains semantic concepts and $E$ contains typed relationships. The planner with parameters $\theta$ produces a retrieval trajectory containing verbatim spans and typed tool calls,

$$T_\theta \sim \pi_\theta(\,\cdot \mid Q, \mathcal{G}, D\,).$$

The trajectory is generated interactively: each action may depend on the deterministic outputs of earlier retrieval calls. The choice and arguments of those calls are stochastic; each invoked retrieval operation is deterministic. Given the trajectory, question, graph, and snapshot, retrieval returns semantic bindings $B$ and a finite route registry $\mathcal{R}$:

$$(B, \mathcal{R}) = \mathrm{Retrieve}(T_\theta, Q; \mathcal{G}, D).$$

For source $s$ and target $t$, the registry contains canonically ordered admissible paths:

$$\mathcal{R}_{s,t}^{(h)} = \mathrm{sort}\left\{ p = (e_1, \ldots, e_k) \in \mathrm{Paths}_{\mathcal{G}}^{\leq h}(s,t) \;\middle|\; \mathrm{governed}(p) \wedge \mathrm{roleValid}(p) \wedge \mathrm{supported}(p) \right\}.$$

Here $h$ is the configured maximum hop count. The finite object is the governed symbolic vocabulary exposed for the question:

$$\Sigma(B, \mathcal{R}) = \mathrm{Subjects}(B) \cup \mathrm{Metrics}(B) \cup \mathrm{Attributes}(B) \cup \mathrm{Operators} \cup \mathrm{IDs}(\mathcal{R}).$$

Every semantic identifier in the submitted plan $P_\theta$ must belong to $\Sigma(B, \mathcal{R})$. Literal payloads are separately obtained from grounded question spans, value probes, or resolved periods and are checked before release. The claim is therefore that the governed symbolic choice set is finite, not that every possible JSON serialization is drawn from a finite language.

The remaining stages are deterministic partial functions:

$$\begin{aligned}
\mathrm{Review}(P_\theta; B, \mathcal{R}) &\to \{\mathrm{accept}, \mathrm{refuse}\}, \\
\mathrm{Compile}(P_\theta; \mathcal{O}, \mathcal{M}) &\to q \in \mathrm{SQL} \;\cup\; \{\perp_{\mathrm{C}}\}, \\
\mathrm{Check}(q; \mathcal{O}, \mathcal{M}, D) &\to F,
\end{aligned}$$

where $\perp_{\mathrm{C}}$ denotes compilation refusal and $F$ is a set of checker findings. The runtime releases SQL only when

$$\mathrm{Review}(P_\theta; B, \mathcal{R}) = \mathrm{accept}, \qquad \mathrm{Compile}(P_\theta; \mathcal{O}, \mathcal{M}) = q, \qquad \nexists f \in F : \mathrm{severity}(f) = \texttt{violation}.$$

The stochastic variables are therefore the retrieval trajectory $T_\theta$ and submitted plan $P_\theta$. Conditional on $T_\theta, Q, \mathcal{O}, \mathcal{M}, D$, retrieval is deterministic. Conditional on the resulting $B, \mathcal{R}$ and a submitted $P_\theta$, review is deterministic. Compilation is additionally deterministic for fixed $\mathcal{O}, \mathcal{M}$, and compiler version; for fixed emitted query $q, \mathcal{O}, \mathcal{M}, D$, checking is deterministic. This contract bounds what the model can name and makes relational realization reproducible. It is not a proof of business correctness: findings classified as `undecidable` are non-blocking, and the planner may select the wrong admissible interpretation.

## 4. Proposed Method

### 4.1 Overview

The evaluated system contains one multi-turn planner agent. It has access only to a fixed typed workflow and has no shell or SQL tool. On each turn, the model either calls one retrieval or review tool or submits a structured plan. Figure 1 shows the workflow and the ownership boundary between planner and engine.

The engine constructs the `submit_plan` JSON schema after resolution. Its enumerations contain only the subjects, metrics, attributes, operators, and route handles retrieved for that question. This prevents an ordinary structured submission from naming an unretrieved object. The engine also validates the submitted plan independently because schema conformance alone does not establish semantic completeness.

### 4.2 The governed inputs

Two authored files constitute the compiler's inputs $\mathcal{O}$ and $\mathcal{M}$: the evaluated ACME governed semantic model is represented as an authored ontology plus a separate physical mapping. The semantic file declares concepts, party roles, edges, and metrics; apart from the free-text `evidence:` justifications shown below, it names no table, column, or join predicate. The mapping file binds each concept and edge to tables, key columns, marker tables, and join predicates. The loader merges the two under eight validation rules that raise rather than warn, so a partially bound model cannot reach the compiler.

An edge declares its endpoints, an optional business role, a cardinality, and a multiplicity classification per direction together with the evidence for that classification (abridged; `description:` fields omitted):

```
edges:
  HOLDS:
    from: Party
    to: Policy
    role: PH                    # policy holder
    cardinality: many_to_many
    fan_out: bounded
    fan_out_reverse: bounded
    evidence: "junction rows per party are small;
              PK(Agreement,Party,Role,Date)"
  SOLD:
    from: Party
    to: Policy
    role: AG                    # selling agent
    cardinality: many_to_many
    fan_out: bounded
    fan_out_reverse: bounded
    evidence: "same junction"
```

`HOLDS` and `SOLD` share endpoints and cardinality and differ only in role, so a role is part of an edge's identity rather than a filter applied to a generic traversal. Every Party-to-Policy edge in ACME is role-bearing (`PH`, `AG`, `UW`), and the longer alternatives run through role objects (`PolicyHolder`, `Agent`, `Underwriter`) whose own mapping carries the same party-role predicate, so no route from Party to Policy arrives without committing a role. The mapping binds each commitment to the predicate the compiler emits (`Party_Role_Code = 'PH'`; §4.7).

Role and kind commitments live in three places, and counting only the first understates the surface: three edges carry a `role:` key, six edges are discriminated by a marker table declared in the mapping (`HAS_LOSS_PAYMENT` restricted to `Loss_Payment`, and likewise for the other amount kinds), and three concepts are role objects with a bound predicate. Nine of the 26 edges and three role concepts therefore encode semantic commitments that the planner can neither introduce nor rewrite.

`fan_out` is one of `none`, `bounded`, or `multiplicative`, declared separately for each direction; ACME's forward declarations are 12 `none`, 13 `bounded`, and one `multiplicative`. The compiler's grain decision reads only whether a traversal can multiply rows: `bounded` and `multiplicative` are the same input to it. Fourteen of the 26 forward declarations are therefore treated as fanning, and the finer distinction is retained for governance review rather than for compilation. These are authored claims about how the business grows rather than measurements. A single observed counterexample can refute a declaration of `none`, whereas finite fixture observations cannot establish the distinction between `bounded` and `multiplicative`. Every ACME edge carries an `evidence` string recording the justification, although the loader permits a declaration without one, and the database probes of §4.4 annotate observed multiplicity without overriding what is declared.

A leaf metric declares its aggregation and the operand it aggregates; composite metrics instead declare a combination of leaf metrics:

```
metrics:
  LossPayment:
    op: sum
    operand: {concept: ClaimAmount, attribute: amount,
              via: HAS_LOSS_PAYMENT}
```

The operand's concept and `via` edge are what let the compiler locate the relation a measure is summed over, and so decide whether the requested output grouping requires the pre-aggregation of §4.7.

### 4.3 Running example

Consider:

> **Q:** What are the total premiums and the total loss of each policy sold by agent 2, by policy number and the policy holder who paid its premiums?

The question combines two measures, two business roles represented through one physical role relation, a long traversal to claim amounts, and grouping at policy-holder grain. It therefore illustrates the decisions that direct SQL generation must otherwise realize implicitly.

### 4.4 Grounding and deterministic retrieval

The planner supplies a decomposition using verbatim spans from the question:

```
subjects:     [Policy, Agent, PolicyHolder]
quantities:   ["total premiums", "total loss"]
attributes:   ["policy number", "policy holder"]
literals:     ["2"]
```

`resolve` maps these spans to governed concepts, metrics, attributes, periods, literals, and candidate routes. The resolver may recover named concepts from the question and reports ambiguity or failure explicitly. It does not ask the model to construct a path.

```
"policy"         → Policy
"total premiums" → PremiumAmount
"total loss"     → TotalLoss
"agent"          → Agent
"policy holder"  → PolicyHolder
"policy number"  → Policy.policy_number
```

The graph enumerator orders governed routes deterministically. Route availability depends on the governed semantic model, physical mapping, and hop bound. Database probes are used separately for literal grounding and diagnostic measured-fan-out annotations; observed multiplicity does not override declared cardinality or make a route admissible.

For multi-hop questions, SPC does not ask the planner to synthesize an edge sequence. The resolver deterministically enumerates routes satisfying governance, role, support, and hop-bound constraints and exposes them as path-opaque handles. The resolver therefore owns route enumeration; when multiple admissible routes remain, the planner retains only the semantic decision of selecting among those returned alternatives. This converts open-ended relational path synthesis into bounded semantic choice among governed alternatives. The representation need not use a particular ontology language; what SPC requires is a governed semantic graph with explicit relationships, roles, route constraints, and grain information sufficient for deterministic traversal and compilation.

```
Input:  grounded spans S, semantic graph G
Output: bindings B, admissible routes for every required target

1  B ← ResolveVocabulary(S, G)
2  for each candidate subject s and required target t:
3      C ← EnumeratePaths(G, s, t, max_hops)
4      𝒫[s,t] ← CanonicalOrder({
           p ∈ C | governed(p) ∧ role_valid(p) ∧ supported(p)
       })
5  report ambiguous and unresolved spans
6  return B, 𝒫
```

### 4.5 Non-composable route handles and bounded plan submission

Routes are returned as deterministically assigned endpoint-plus-ordinal handles such as `Policy>PolicyAmount#4`. The handle identifies a selectable route without encoding an edge chain that the model could edit or combine. When several routes connect the same endpoints, the planner can inspect their edge sequences and fan-out annotations before selecting one. Because the submission schema is constructed from the question-specific resolution result, an unretrieved route handle is ordinarily unrepresentable; independently, the engine rejects invented or composed handles.

The submitted plan contains semantic objects and handles, not physical SQL objects:

```json
{"subject":   "Policy",
 "measures":  [{"metric": "PremiumAmount", "route": "Policy>PolicyAmount#4"},
               {"metric": "TotalLoss",     "route": "Policy>Claim#1"}],
 "dimensions":[{"attribute": "policy_number", "route": "SELF"},
               {"attribute": "id", "route": "Policy>PolicyHolder#1"}],
 "filters":   [{"attribute": "id", "route": "Policy>Agent#1",
                "operator": "=", "value": "2"}]}
```

For example, `describe_routes` can show that two handles reaching `PolicyAmount` have different traversals:

```
Policy>PolicyAmount#4   COVERED_BY → PRICED_PREMIUM
Policy>PolicyAmount#6   POLICY_LOCATED_AT → … → PRICED_PREMIUM
```

The model chooses between these interpretations after inspecting their descriptions, but the graph and physical joins remain engine-owned.

### 4.6 Deterministic review and retry

`review_plan` compares a draft against the resolution result. It checks, among other conditions, that retrieved required metrics and attributes are not silently omitted and that independently fanning-out branches are not projected as though they formed one row. The skill requires the planner to call this tool before submission. The engine then re-runs the same review on the exact submitted plan; an error finding refuses the attempt.

A refused attempt is not edited in place. With the evaluated default configuration, the engine permits one fresh retry. The earlier conversation is discarded and only the last refusal reason is appended to the new system prompt. This retry policy is part of the evaluated SPC system.

### 4.7 Grain-aware SQL compilation

After review, route handles are resolved to governed semantic paths and physical joins. Role predicates are derived from typed edges; for the running example, the Agent route introduces the `AG` predicate and the PolicyHolder route introduces the `PH` predicate. The planner cannot omit or rewrite those predicates.

The compiler uses a spine/satellite layout for supported grain changes. The spine contains one row per requested output group. Each measure is computed in its own satellite at the measure's supported source grain, and the aggregated satellites are then joined back to the spine. This prevents independent measure rows from multiplying one another before aggregation. SQL is constructed as a `sqlglot` abstract syntax tree and rendered for SQLite.

```sql
WITH __spine AS (
  SELECT DISTINCT ... FROM Policy
  INNER JOIN Agreement_Party_Role
    ON ... AND Party_Role_Code = 'PH'
), __g0 AS (
  SELECT DISTINCT ...
  INNER JOIN Agreement_Party_Role
    ON ... AND Party_Role_Code = 'AG'
  ...
), __m0 AS (
  SELECT __k0, __g0, SUM(...) AS __v FROM __g0 ... GROUP BY ...
), ...
SELECT __spine.policy_number, __spine.id,
       SUM(__m0.__v + ...) AS PremiumAmount,
       SUM(__m1.__v + ...) AS TotalLoss
FROM __spine JOIN __m0 ... JOIN __m3 ...
GROUP BY __spine.policy_number, __spine.id;
```

```
Input:  canonical plan P, ontology O, physical mapping M
Output: SQL, or REFUSE

1  resolve each semantic path to physical joins via M
2  derive role predicates from typed edges
3  output_grain ← group keys of P
4  for each measure μ in P:
       measure_grain ← declared grain of μ's operand
       if measure_grain = output_grain:      lower μ inline
       else if supported transform exists:   lower μ through a
                                             pre-aggregated satellite
       else:                                 REFUSE
5  derive GROUP BY, DISTINCT, aliases
6  construct and render the SQL AST
7  return SQL
```

### 4.8 Static checks and their limit

The checker parses emitted SQL and evaluates declared-join coverage, role predicates, recognizable metric definitions, literal grounding, and supported grain patterns. A finding is a `violation` when a checker rule deterministically establishes that a declared governed condition is violated. It is `undecidable` when the query falls outside the checker's supported analysis. The evaluated runtime blocks `violation` findings but does not block `undecidable` findings. Accordingly, the paper uses *checked* rather than *formally certified* for the end-to-end result.

The deterministic guarantee is conditional: for a fixed plan, ontology, mapping, and compiler version, the compiler emits the same SQL. It does not guarantee that the planner selected the business-correct plan or that the static checker is complete.

### 4.9 Why counterfactual data are required

On a small fixture, a naive join and a grain-safe lowering may return the same number. A counterfactual instance can alter an irrelevant relationship multiplicity while preserving the target business fact:

| Query form | Shipped instance | Counterfactual instance |
|---|---:|---:|
| Naive join | 2 | 3 (inflated) |
| Grain-safe lowering | 2 | 2 |
| Reference | 2 | 2 |

This form of data instrumentation tests whether the fixture distinguishes the specific wrong path or grain transformation under study. It cannot establish that the authored reference is business-correct; external adjudication is a separate requirement.

## 5. Experimental Design

### 5.1 Dataset and question classes

We use the enterprise insurance benchmark introduced by Sequeda, Allemang, and Jacob [Sequeda], using the ACME adaptation employed by dbt [dbt-2026] and using a SQLite port of the database and execution harness while retaining the source reference SQL except for a recorded lexical comment normalization. In the archived dbt benchmark artifact [dbt-acme-artifact], the investigation enumerates 44 inquiry identifiers but materializes 43 as `QandA:Inquiry` resources. The study's archived Turtle question artifact is a byte-identical copy of the original data.world benchmark artifact [acme-source-artifact], which materializes the listed `IQ_d51d706e4b7ef001706289b940f09b24` item ("claims closed in 2019") with its SQL and SPARQL references. The study copy and the original data.world source artifact both have SHA-256 `a29a10dad7666723ccdd91831e7261cffd0847a567a1bf2dee612d7fa7844a34`. Thus the item was recovered from the original public source rather than reconstructed by the study. The study loader traverses each inquiry to its SQL references and, when alternatives exist, tries candidates in fixed query-ID-sorted order and selects the first that executes successfully against the SQLite benchmark database. An artifact audit confirmed the 44-listed/43-materialized dbt artifact and the byte identity of the 44-item study artifact with the data.world source. This local universe is distinct from the 11-question subset reported in dbt's published comparison [dbt-2026]. The questions include single-concept lookups, multi-hop traversals, role-sensitive relationships, aggregation at multiple grains, and combined requests. These labels are study-specific descriptions, not benchmark-provided categories.

### 5.2 Counterfactual multiplicity evaluation

**Definition.** A counterfactual database D′ is *discriminating* for query q if it preserves the business facts relevant to q's intended measure while changing relational multiplicities that q's semantics does not pin, such that a grain-unsafe implementation of q diverges from a grain-correct reference on D′ while agreeing with it on the shipped instance D. Each counterfactual is specified together with the independent reference it must preserve, so the perturbation's semantic validity is checkable.

The counterfactuals are targeted tests of enumerated failure modes, not independent samples from a population of enterprise databases. Their evidential role is to show that a proposed wrong path or grain transformation is separable from the reference on at least one semantics-preserving instance.

### 5.3 Arms

The paired comparison contains the two systems shown in Figure 2.

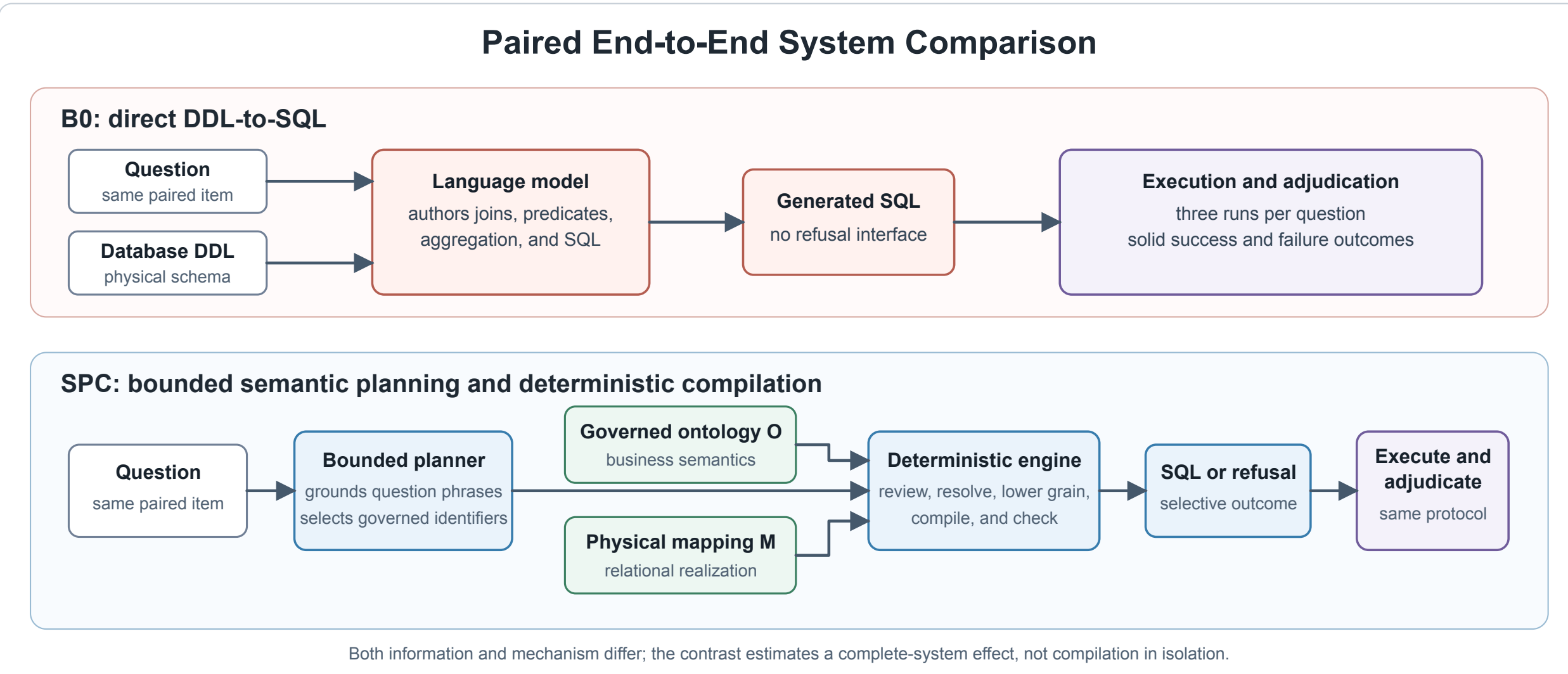

**Figure 2.** Experimental arms. B0 generates SQL directly from the question and DDL. SPC selects a governed semantic plan and delegates relational realization to deterministic software. The comparison therefore measures complete systems rather than compilation in isolation.

**B0** receives the benchmark DDL and question and generates SQL directly. It has no ontology, compiler, or refusal mechanism. **SPC** receives the authored semantic model and physical mapping through its typed workflow. Because information and mechanism both differ, the contrast estimates an end-to-end system effect. dbt's published numbers are contextual and are not pooled with either arm.

### 5.4 Protocol and metrics

The primary paired comparison uses `gpt-5.6-luna` and three runs per question. B0's run metadata records a direct first-party endpoint, Python 3.12.13, and hashes of its question source, DDL, database, ontology, and mapping. The harness requested temperature 0.0, but the provider rejected that parameter and B0 records `temperature_sent: provider default`. The SPC summary artifact does not preserve transmitted-temperature or endpoint metadata. Accordingly, this paper does not claim verified sampling-parameter parity, even though both campaign scripts requested 0.0.

Each SPC run permits at most one fresh retry after a deterministic refusal. B0 makes one direct generation call and does not refuse. This difference is part of the systems comparison rather than a controlled component ablation. On the 114 primary paired outcomes, 10 runs invoked the retry; nine retries produced SQL and one remained a refusal. The SPC arm therefore used 124 planner attempts for 114 reported run outcomes. A planner attempt may contain multiple model turns; total model-call, token, and cost parity with B0 was not measured.

Two SPC robustness boards use the same workflow, adjudicated set, and three-run protocol with `gpt-5.4` and `gemini-3.6-flash`. They are descriptive model-sensitivity results; no secondary-model B0 board is available, so they are not included in the paired hypothesis test.

| Board | Preserved generation provenance |
| --- | --- |
| Luna B0 | Direct first-party API; three runs; requested temperature 0.0 rejected and provider default transmitted |
| Luna SPC | Typed-tool workflow; three runs; requested temperature 0.0; transmitted temperature and endpoint not preserved in the aggregate |
| GPT-5.4 SPC | Same typed-tool workflow and three-run protocol; final per-request transport metadata not preserved |
| Gemini-3.6-Flash SPC | Same typed-tool workflow and three-run protocol; final per-request transport metadata not preserved |

Correctness is determined by executing prediction and reference SQL and then classifying disagreements. Deterministic comparison settles identical same-width row-value multisets. Projection differences are detected and characterized deterministically, but their semantic verdict and all other unsettled cases are decided by `gpt-4o` using the question, both SQL programs, and both executed results, with up to three votes and a majority rule. The judge record preserves the `gpt-4o` alias but not the exact model snapshot. The verdict set is `equivalent`, `superset`, `prediction_wrong`, `agrees_on_this_data_only`, `gold_defective`, and `ambiguous_question`. The frozen scoring rule counts `equivalent` and `superset` as success for both arms and discloses supersets separately. Because that owner ruling is debatable, we also report a strict sensitivity analysis in which only `equivalent` counts. The judge may assign `agrees_on_this_data_only` when the SQL structures indicate a latent semantic difference despite agreement on the available execution evidence. These flags are not counterfactual confirmations; they are reported separately and tested under a conservative sensitivity that counts them as successes.

### 5.5 Comparison set and estimands

The primary paired set contains 38 questions. Six items are removed uniformly from both arms under a fixed registry:

| Failure class | Items |
| --- | --- |
| Reference fan trap under admissible multiplicities | 2 |
| SQLite arithmetic incompatibility in source references | 2 |
| SQLite-incompatible source function (`DATE_DIFF`) | 1 |
| Adjudicated over-projection / question-scope dispute | 1 |

The registry was constructed post hoc during reference auditing after system outputs had been inspected, then frozen before all arms were rescored and the final comparison tables were regenerated. It is not a preregistered exclusion set.

The over-projection item was equivalent for SPC on all three runs but is removed because the registry is applied uniformly, not because SPC failed it. Its removal is conservative for the arm difference: B0 was not solid on that item, so retaining it would add another SPC-only discordance.

The all-item analysis retains all 44 questions and counts missing or unsuccessful items against the system. The B0 raw artifact contains 43 questions because the dialect-incompatible item was not run; it is therefore counted as unsuccessful on the conservative 44-item denominator.

The primary estimand is **solid success** on the 38 paired questions: a question succeeds only if all three runs are accepted. We also report at-least-once success and run-level outcome classes. The paired hypothesis test is a two-sided exact McNemar test over question-level solid success. Wilson 95% intervals describe individual proportions but are not used as the paired significance test. The McNemar result characterizes discordance on this fixed paired benchmark; because the questions are not a random sample of enterprise workloads, it does not establish cross-domain generalization.

### 5.6 Reproducibility scope

The primary and sensitivity statistics were regenerated from frozen verdict artifacts with the fixed registry applied uniformly. The secondary boards are the final prepared three-run summaries, but their per-run verdict files are not archived. The execution database, counterfactual environment, and exact working-tree state of the primary campaign are also unavailable. Consequently, the primary aggregate statistics are auditable, but the study is not currently reproducible end to end.

## 6. Results

### 6.1 Primary paired comparison

Table 1 reports the frozen inclusive scoring rule on the 38-question paired set.

| Outcome | B0: DDL→SQL | SPC |
|---|---:|---:|
| Solid success | 21/38, 55.3% [39.7, 69.9] | **37/38, 97.4% [86.5, 99.5]** |
| At least one successful run | 31/38, 81.6% [66.6, 90.8] | **38/38, 100% [90.8, 100]** |
| Successful runs | 78/114 | **113/114** |
| Refusals | 0/114 | 1/114 |
| Adjudicated wrong runs | 29/114 | **0/114** |
| Data-only divergence flags | 7/114 | **0/114** |

B0's 78 successful runs comprise 70 `equivalent` and eight `superset` verdicts. It also produced 29 `prediction_wrong` verdicts and seven `agrees_on_this_data_only` flags. The latter are judge-inferred structural risks, not mechanically confirmed counterfactual failures, and are therefore not merged with the adjudicated-wrong count. SPC produced 113 `equivalent` verdicts and one `no_sql` refusal. These counts show a change in failure character as well as aggregate success: on this comparison set, the direct generator never refused and returned 29 adjudicated wrong results that still executed, whereas SPC's only unsuccessful attempt returned no SQL.

The paired solid-success table is:

| | SPC solid | SPC not solid |
|---|---:|---:|
| B0 solid | 21 | 0 |
| B0 not solid | 16 | 1 |

The two-sided exact McNemar test on the 16-to-0 discordance gives $p = 3.0518 \times 10^{-5}$. The estimated difference in solid success is 42.1 percentage points. The Wilson intervals describe each arm separately; the McNemar result is the inferential comparison because the same questions appear in both arms. It applies to these 38 paired benchmark questions rather than to a random population of enterprise queries.

### 6.2 Sensitivity to the superset ruling

Under strict scoring, where only `equivalent` counts as success, B0 solid success falls from 21/38 to 18/38 (47.4% [32.5, 62.7]); SPC remains 37/38 (97.4% [86.5, 99.5]). At-least-once success is 29/38 (76.3%) for B0 and 38/38 for SPC. The paired table contains 19 SPC-only solid questions and no B0-only solid question, giving a two-sided exact McNemar $p = 3.8147 \times 10^{-6}$.

The main conclusion therefore does not depend on counting over-projected supersets as correct. The inclusive rule is retained as the frozen owner ruling, while the strict analysis exposes its effect.

A second conservative sensitivity treats every `agrees_on_this_data_only` flag as a success rather than as evidence of latent divergence. B0 then rises to 22/38 solid questions, while SPC remains 37/38. The paired table contains 15 SPC-only solid questions and no B0-only solid question, yielding an exact two-sided McNemar $p = 6.1035 \times 10^{-5}$. The primary conclusion is therefore not dependent on accepting the judge's structural interpretation of those seven runs.

### 6.3 Model robustness

The same SPC workflow and three-run protocol were evaluated with two additional planner models on the 38-question adjudicated set.

| SPC planner | Solid success |
|---|---|
| `gpt-5.6-luna` | **37/38, 97.4%** [86.5, 99.5] |
| `gpt-5.4` | 36/38, 94.7% [82.7, 98.5] |
| `gemini-3.6-flash` | 35/38, 92.1% [79.2, 97.3] |

All three models remain within two solid questions of one another. These boards support robustness to the tested planner choice but do not isolate model effects because only Luna has a paired B0 comparison.

### 6.4 Conservative all-item accounting

Table 2 retains every current benchmark item, including the six registry items and B0's unrun dialect-incompatible question.

| Outcome on 44 items | B0: DDL→SQL | SPC |
|---|---|---|
| Solid success | 22/44, 50.0% [35.8, 64.2] | **38/44, 86.4% [73.3, 93.6]** |
| At least one successful run | 35/44, 79.5% [65.5, 88.8] | **39/44, 88.6% [76.0, 95.0]** |

This table is deliberately conservative but is not a clean estimate of system correctness: it mixes model failures with reference-semantic faults and a dialect failure. Its purpose is to show the effect of the adjudication registry rather than allow exclusions to disappear from reporting.

The corresponding all-44 paired solid-success table contains 21 questions solid for both systems, 17 solid only for SPC, one solid only for B0, and five solid for neither. The two-sided exact McNemar test on the 17-to-1 discordance gives $p = 1.4496 \times 10^{-4}$. This sensitivity removes dependence on the post-hoc 38-item denominator, although it is not a clean correctness estimate because it deliberately retains reference-semantic and execution-port incompatibilities.

### 6.5 Counterfactual and checker evidence

The stored counterfactual tests contain cases in which a grain-unsafe query and the reference agree on the shipped instance but diverge after an additional role or fan-out row is introduced while the target business quantity is held fixed. The checker evaluation covers nine deliberately unsafe mutations and four corresponding grain-safe controls. Eight of the nine unsafe forms change the executed result on the counterfactual instance; all nine are flagged, and the four safe controls preserve the result and are not flagged.

This evidence supports the narrower claim that the counterfactual instances discriminate the enumerated multiplicity defects. It does not show that all semantically wrong queries are separated, nor does it validate the business meaning of the authored references. Moreover, the database files needed to re-run these tests are absent from the current archive (§5.6), so this part of the evidence is presently auditable in code and documentation but not re-executable from the archive.

### 6.6 Interpretation

For a fixed accepted plan and fixed artifacts, relational realization is deterministic. End-to-end behavior is not: one of 114 SPC run outcomes refused, and the three-run horizon is too short to estimate long-run stability. The result should therefore be read as relocation of stochasticity from SQL construction to bounded semantic planning, not elimination of model variance.

The gain also comes with substantial authored structure and implementation cost. The study does not determine how much of the difference is due to the semantic knowledge, the bounded interface, deterministic compilation, the retry policy, or their interaction. It establishes that the complete SPC system outperformed the complete DDL baseline on the paired ACME set.

## 7. Discussion

The compiler, resolver, and checker were developed while inspecting this benchmark. Although their implementation is organized around reusable relational constructs rather than question identifiers, this repository does not contain a frozen held-out evaluation of unseen construct combinations. Accordingly, the paper makes no empirical claim that the measured accuracy generalizes beyond ACME or that implementation effort grows only by construct rather than by workload. A held-out composition study or a second enterprise domain is future work, not part of the reported result.

Within this scope, the defining design is not deterministic SQL rendering by itself. It is the combination of (i) a finite question-specific semantic choice set, (ii) engine-owned relational realization, and (iii) explicit refusal paths. The experiment cannot estimate the contribution of each component, but the run-level outcomes show why the combination matters operationally: an executable wrong result and a refusal are not interchangeable.

That same bound limits coverage. If the intended business composition is not represented by an admissible route, SPC cannot infer or approve it merely because the physical tables are connected. It must refuse the query until the semantic model or physical mapping is extended.

The counterfactual analysis also clarifies the role of test data. Static rules can name known hazards, but a benchmark instance must contain multiplicities that make those hazards observable. Conversely, discriminating data establish only that the reference is separable from enumerated alternatives; they do not establish that the reference captures the intended business meaning.

## 8. Practicality

The proposed architecture moves substantial cost to development time: governed semantic-model authoring, physical mappings, compiler and checker implementation, regression testing, and ongoing semantic governance. It should not be interpreted as a universal replacement for conventional semantic-layer or text-to-SQL systems.

For ad hoc databases, lightly modeled domains, and predominantly simple analytical questions, semantic context combined with direct LLM SQL generation may provide a better engineering trade-off. It requires less up-front modeling and can accommodate previously unseen query structures without extending a compiler.

Deterministic semantic compilation becomes more attractive when three conditions coincide:

1. business semantics are stable and reused across many queries;
2. questions frequently involve deep relational paths, role-sensitive relationships, aggregation-grain changes, or other high-risk constructs;
3. incorrect but plausible answers carry sufficient operational cost that repeatability and explicit refusal are valued over unrestricted query flexibility.

The architecture therefore trades **modeling and compiler complexity at development time** for **reduced stochastic relational reasoning at query time**.

An important practical question is whether compiler complexity grows by reusable relational construct or by individual benchmark question. The intended design is the former: filters, joins, aggregation, grain reconciliation, temporal predicates, ordering, and related constructs are implemented as generic lowering rules. The held-out evaluation described in §7 would be needed to test this distinction directly; it is not reported here.

The expected deployment model is consequently hybrid rather than universal. Queries expressible within the governed semantic model can use deterministic compilation, while unsupported semantics may be clarified, refused, or handled by a less constrained fallback depending on the application's risk requirements.

## 9. Threats to Validity

**Single domain.** The evaluation uses the ACME Insurance domain. Insurance provides the multi-hop, role-sensitive, and multiplicity-sensitive structures required by the study, but results should not be interpreted as evidence that the same effects hold across all enterprise domains.

**Limited model diversity.** SPC is evaluated with three planner configurations across two model families, but only Luna has a paired baseline. The experiment therefore does not establish broad invariance across providers, model families, or future model versions.

**Public benchmark exposure.** The benchmark predates the evaluated model and may have appeared in training data. Because both arms use the same named model, such exposure is partly controlled in the paired contrast, but absolute success rates may still be inflated.

**Researcher involvement.** The authors developed the system, ontology, compiler, counterfactuals, and much of the evaluation infrastructure. This creates implementation and adjudication bias that cannot be eliminated by internal checks. The principal mitigations are frozen verdict artifacts, uniform application of the six-item registry, independent reference computations for documented semantic defects, strict-scoring sensitivity, and reporting the original denominator.

**Development on the benchmark.** Compiler, resolver, and guardrail improvements were made while examining benchmark questions. This creates a risk that generic-looking rules encode benchmark-specific behavior. No held-out construct-combination result is reported, so no cross-workload generality claim is made (§7).

**Counterfactual validity.** Counterfactual multiplicity evaluation is meaningful only if each perturbation preserves the business fact required by the natural-language question while changing multiplicities irrelevant to that fact. The current counterfactuals enumerate hazards chosen by the authors and cannot detect wrong paths that were not anticipated.

**Reference-answer adjudication.** Some published benchmark answers are removed from the paired comparison. Two contain fan traps supported by semantic and counterfactual evidence independent of SQLite. Two source ratio queries execute under SQLite but acquire unintended integer-division semantics, and one uses the unsupported BigQuery `DATE_DIFF` function; these three are execution-port incompatibilities rather than claims that the source queries are defective in their native dialect. One item depends on an owner ruling about over-projection. The registry was constructed post hoc, and the last category is particularly subjective. The 44-item analysis and strict sensitivity prevent these decisions from being hidden.

**LLM-based judging.** The headline success measure is adjudicated rather than pure execution match. Although the judge sees both SQL programs and executed rows, uses deterministic short-circuits, votes up to three times, and defaults to `prediction_wrong` on failure, it is still an authored and model-dependent instrument. In particular, `agrees_on_this_data_only` is a structural judgment rather than a mechanically demonstrated counterfactual failure. The five-case calibration described in the repository is small, and the nontrivial verdict classes have not been independently adjudicated by a second human reviewer. The conservative data-only sensitivity in §6.2 bounds the effect of that specific category but does not replace independent adjudication.

**Semantic-model authoring advantage.** The proposed system receives structured enterprise knowledge that may not be available to raw-schema baselines. The B0-vs-SPC comparison therefore estimates the end-to-end system effect and cannot separate gains due to additional semantic knowledge from gains due to bounded planning or deterministic compilation. The paper does not make that causal attribution.

**Unequal execution protocols.** SPC may make one fresh retry after a deterministic refusal, whereas B0 makes one direct generation call and has no refusal interface. This is intentional in the end-to-end comparison but prevents attributing the difference to the compiler alone. Ten of 114 primary SPC outcomes invoked the retry, nine of which produced SQL.

**Sampling provenance.** B0 records that the requested temperature was dropped and the provider default was used. The SPC aggregate lacks the equivalent transmitted-temperature and endpoint record. Named model and requested temperature are therefore known, but exact sampling parity is not. The judge is recorded only as the mutable `gpt-4o` alias; its exact snapshot was not preserved.

**Checker incompleteness.** The SQL checker blocks deterministically established violations but the evaluated runtime does not block findings marked `undecidable`. A checked query is not a formally certified query, and zero observed wrong-but-executed SPC runs does not establish a general soundness guarantee.

**Artifact completeness.** The current archive lacks the SQLite databases and an environment sufficient to re-run the benchmark. Headline statistics can be regenerated from frozen verdict files, but the execution evidence and counterfactual suite cannot presently be reproduced end to end. The secondary-model boards are the final prepared three-run results on the same adjudicated dataset; their per-run verdict files are not archived. The primary SPC trace records an uncommitted working-tree state, but the corresponding patch was not archived, so that evaluated code state is also not exactly reconstructible.

## 10. Conclusion

This study evaluates a text-to-SQL architecture in which a language model selects governed semantic objects but does not author relational structure or SQL. On the 38-question paired ACME set, SPC achieved solid success on 37 questions, compared with 21 for direct DDL-to-SQL generation, and produced no wrong-but-executed run in 114 run outcomes. The strict-scoring sensitivity strengthened rather than weakened the difference. Under the same three-run SPC protocol, GPT-5.4 was solid on 36/38 questions and Gemini-3.6-Flash on 35/38.

The result is an end-to-end systems result on one benchmark. It does not prove that deterministic compilation alone caused the gain, that the checker is sound for arbitrary SQL, or that the result transfers across domains and models. What it does show is that, for this governed workload, an SPC system that moves path, role, grain, and SQL realization into explicit deterministic machinery achieved substantially higher repeated-run reliability while preserving answer coverage.

## Code and Data Availability

A reference implementation of the semantic planner, path resolver, grain-aware compiler, and static checker, together with the ACME semantic model and its physical mapping, is available under the MIT license at `github.com/yai333/Semantic-Planning-and-Deterministic-for-Text-to-SQL`. That snapshot is deliberately partial: it excludes the benchmark questions, gold labels, database fixtures, evaluation harness, frozen verdict artifacts, and execution traces. This preprint is archived at `doi.org/10.5281/zenodo.21966805`.

The reproducibility scope of §5.6 and §9 is therefore the operative one, and the public snapshot narrows it further: the reported aggregate statistics were regenerated from frozen verdict files that are not part of that snapshot, and the execution databases and exact working-tree state of the primary campaign are not published. The published compiler and checker implementations can be inspected and exercised from the public code snapshot; the reported measurements cannot presently be reproduced end to end from public artifacts alone.

## Authoring and Tool Disclosure

Generative AI tools were used during software development and manuscript editing. The authors remain responsible for the experimental design, code, artifact provenance, literature verification, statistical analysis, and all claims in the paper.

## References

[Allemang-2024] Allemang, D., and Sequeda, J. *Increasing the LLM Accuracy for Question Answering: Ontologies to the Rescue!* arXiv:2405.11706, 2024.

[dbt-2026] Ganz, J., and Perigaud, B. *Semantic Layer vs. Text-to-SQL: 2026 Benchmark Update.* dbt Developer Blog, 7 Apr. 2026.

[dbt-join-logic] dbt Labs. *Joins.* dbt Developer Hub, accessed 16 Aug. 2026.

[dbt-acme-artifact] dbt Labs. *ACME benchmark questions artifact* (`benchmark_questions.ttl`). dbt LLM Semantic Layer Benchmark repository, commit `a29f2429`, accessed 17 Aug. 2026. https://github.com/dbt-labs/dbt-llm-sl-bench/blob/a29f2429b1bb38cee9d591892299e96f1e197ff9/benchmark_questions.ttl.

[acme-source-artifact] data.world. *ACME Insurance benchmark artifact* (`ACME_Insurance/investigation/acme-benchmark.ttl`). data.world CWD Benchmark Data repository, commit `0b75eb62`, accessed 17 Aug. 2026. https://github.com/datadotworld/cwd-benchmark-data/blob/0b75eb62eaf7ea315a863cd7611ebc908149f7e0/ACME_Insurance/investigation/acme-benchmark.ttl.

[Guo-2019] Guo, J., Zhan, Z., Gao, Y., Xiao, Y., Lou, J.-G., Liu, T., and Zhang, D. *Towards Complex Text-to-SQL in Cross-Domain Database with Intermediate Representation.* Proceedings of the 57th Annual Meeting of the Association for Computational Linguistics, pp. 4524–4535, 2019. doi:10.18653/v1/P19-1444.

[Hurtado-2001] Hurtado, C. A., and Mendelzon, A. O. *Reasoning about Summarizability in Heterogeneous Multidimensional Schemas.* International Conference on Database Theory, pp. 375–389, 2001. doi:10.1007/3-540-44503-X_24.

[Karayannidis-2026] Karayannidis, N. *Grain Theory: Type-Level Granularity Correctness in Data Pipelines.* arXiv:2601.00995, version 2, 2026.

[Lenz-1997] Lenz, H.-J., and Shoshani, A. *Summarizability in OLAP and Statistical Data Bases.* Proceedings of SSDBM, pp. 132–143, 1997.

[Ontop-2017] Calvanese, D., Cogrel, B., Komla-Ebri, S., Kontchakov, R., Lanti, D., Rezk, M., Rodriguez-Muro, M., and Xiao, G. *Ontop: Answering SPARQL Queries over Relational Databases.* Semantic Web 8(3):471–487, 2017.

[Scholak-2021] Scholak, T., Schucher, N., and Bahdanau, D. *PICARD: Parsing Incrementally for Constrained Auto-Regressive Decoding from Language Models.* Proceedings of EMNLP, pp. 9895–9901, 2021. doi:10.18653/v1/2021.emnlp-main.779.

[2604.25149] Rumiantsau, M., and Fokeev, I. *Semantic Layers for Reliable LLM-Powered Data Analytics: A Paired Benchmark of Accuracy and Hallucination Across Three Frontier Models.* arXiv:2604.25149, 2026.

[2606.31041] Kim, H. J., Khoeurn, S., and Yoon, Y. J. *A Semantic-Layer-Mediated Agent for Natural Language to SQL over Heterogeneous Enterprise Databases.* arXiv:2606.31041, 2026.

[2604.06736] Zhou, Y., Zhang, F., Guo, Z., Chen, Y., Zhang, H., Nakov, P., and Xie, Z. *SQLStructEval: Structural Evaluation of LLM Text-to-SQL Generation.* arXiv:2604.06736, 2026.

[Saha-2016] Saha, D., Floratou, A., Sankaranarayanan, K., Minhas, U. F., Mittal, A. R., and Özcan, F. *ATHENA: An Ontology-Driven System for Natural Language Querying over Relational Data Stores.* Proceedings of the VLDB Endowment 9(12):1209–1220, 2016.

[Sen-2020] Sen, J., Lei, C., Quamar, A., Özcan, F., Efthymiou, V., Dalmia, A., Stager, G., Mittal, A. R., Saha, D., and Sankaranarayanan, K. *ATHENA++: Natural Language Querying for Complex Nested SQL Queries.* Proceedings of the VLDB Endowment 13(11):2747–2759, 2020.

[Sequeda] Sequeda, J., Allemang, D., and Jacob, B. *A Benchmark to Understand the Role of Knowledge Graphs on Large Language Model's Accuracy for Question Answering on Enterprise SQL Databases.* arXiv:2311.07509, 2023.

[TrustSQL] Lee, G., Chay, W., Cho, S., and Choi, E. *TrustSQL: Benchmarking Text-to-SQL Reliability with Penalty-Based Scoring.* arXiv:2403.15879, 2024.

[Yan-1995] Yan, W. P., and Larson, P.-Å. *Eager Aggregation and Lazy Aggregation.* Proceedings of VLDB, pp. 345–357, 1995.

[Zhong-2020] Zhong, R., Yu, T., and Klein, D. *Semantic Evaluation for Text-to-SQL with Distilled Test Suites.* Proceedings of EMNLP, pp. 396–411, 2020. doi:10.18653/v1/2020.emnlp-main.29.